\documentclass[twocolumn,amsmath,amssymb]{revtex4}

\usepackage{graphicx}
\usepackage{dcolumn}
\usepackage{bm}
\begin{document}


\title{Demonstration of high-$Q$ TE-TM photonic crystal nanobeam cavities}

\author{Murray W. McCutcheon, Parag B. Deotare, Yinan Zhang and Marko Lon\v{c}ar}
\affiliation{School of Engineering and Applied Sciences, Harvard University, Cambridge, MA 02138
}

\date{\today}

\begin{abstract}

We experimentally demonstrate high Quality factor dual-polarized TE-TM photonic crystal nanobeam 
cavities.  The free-standing nanobeams are fabricated in a 500 nm thick silicon layer, and
are probed using both tapered optical fiber and free-space resonant scattering set-ups.  We 
measure $Q$ factors greater than $10^4$ for both TM and TE modes, and observe large 
fiber transmission drops (0.3 -- 0.4) at the TM mode resonances.

\end{abstract}

\maketitle

Recently there has been significant interest in nanobeam photonic crystal microcavities.  
Although first proposed more than a decade ago~\cite{Foresi}, it was only recently shown 
that this architecture was 
capable of supporting ultra-high {\em Q}-factor cavities with dimensions on the order of a 
cubic wavelength in material~\cite{Notomi_1D, McCutcheon_08, Chan_09}.  Since then, there have 
been a host of new applications proposed and demonstrated, including optomechanical 
crystals~\cite{Eichenfield_09b}, visible nanocavities~\cite{Gong, Khan, Murshidy},
low-power reconfigurable switches~\cite{Frank_10}, biosensors~\cite{DiFalco_09, Wang_10},
high transmission waveguides~\cite{Quan_10, Zain_08b} and lasers~\cite{Halioua_10,Zhang_10}. 
Recently, we showed theoretically that nanobeams could be engineered to have both TE 
and TM stopbands, provided the nanobeam is thick enough to support TM guided modes~\cite{Zhang_09}.  
High Q-factor cavity modes can be designed for both polarizations {\em simultaneously} by using
an appropriate lattice tapering.  By tuning the aspect ratio of the nanobeam (namely the height 
and width), the energy separation of the modes can be tuned.  Dual-polarized nanocavities have 
the potential to open up new applications in chip-scale nonlinear optics.  In particular, we 
have recently proposed such a system for single-photon frequency conversion in III-V 
materials~\cite{McCutcheon_OE09, Burgess_09b}.  These devices also raise the 
intriguing prospect of photonic crystal quantum cascade lasers~\cite{Wakayama_08, Loncar_QCL}, 
for which the radiation is TM-polarized.  

\begin{figure}[htb]
\begin{center}
\includegraphics[width=7.5cm]{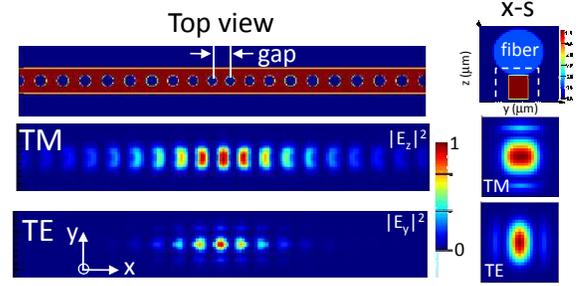}
\end{center}
\vspace{-10pt}
\caption{The field intensity profiles $|E_z|^2$ ($|E_y|^2$) for the TM (TE) modes are shown
from the top and in cross-section (x-s).  The x-s fields were calculated with a 1 micron diameter 
fiber in contact with the top of the cavity, and are plotted over the white dashed square shown
in the index profile.  The larger mode volume of the TM mode is clearly visible from its greater 
lateral extent, and there is significantly greater overlap of the TM mode with the fiber compared
to the TE mode.
\label{fig:modes}}
\end{figure}

In this letter, we report the experimental observation of TE/TM dual-mode cavities in silicon. 
Our cavity design exploits the principles we (and others) have outlined in 
prior work~\cite{Lalanne_JQE, Sauvan, McCutcheon_08,Chan_09,Zhang_09,Deotare_09}.  
We have shown that in principle, nanobeam cavities can be designed to have both TE and TM 
modes with Quality factors exceeding $10^6$, but this requires a very thick structure 
with thickness:width:period ratio of 3:1:1~\cite{Zhang_09}.  Here, given our silicon 
device layer thickness 
of 500 nm and our operating wavelength near 1500 nm, the cavities were designed with TE and TM modes 
with Q factors of $7 \times 10^6$ and 120,000, and mode volumes of 0.56 and 1.37, respectively.
The nominal designs have nanobeam widths of 380-400 nm, hole pitch $a = 330$ nm, radius $r/a =
0.265$, and a symmetric 6-period taper down to a pitch of $0.84a$ in the center of the cavity.  
The TE and TM modes are 
separated by 50 nm, with the TE mode at higher energy.  The field intensity profiles for 
the modes extracted from 3D finite-difference time-domain (FDTD) simulations are shown in 
Figure~\ref{fig:modes}.    

\begin{figure}[t]
\begin{center}
\includegraphics[width=7.5cm]{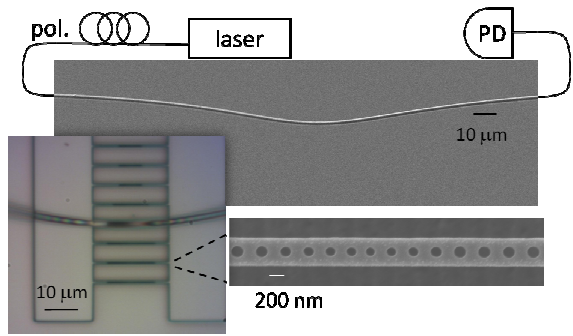}
\end{center}
\vspace{-10pt}
\caption{SEM image of a dimpled fiber taper and a schematic of the optical set-up 
(PD - photodiode).  The optical image (left) shows the fiber in direct contact with a
 nanobeam cavity, which is shown in greater detail in the SEM inset.  
\label{fig:fiber}}
\end{figure}

Because TM modes have the dominant electric field component oriented orthogonal to the device plane, 
they must be excited via evanescent~\cite{Srin_PRB04} 
or end-fire waveguide coupling techniques~\cite{Notomi_04}. They cannot be 
readily probed from free-space~\cite{McCutcheon_05}, since like an electric dipole, 
they do not radiate parallel to
the axis of the dipole moment.  In order to probe our cavities, we use two complimentary techniques.
Firstly, we employ a tapered fiber optical set-up.  We pull an SMF-28 telecom fiber heated with a 
hydrogen torch to a diameter close to 1 $\mu$m.  The fiber is
mounted in a U-shape, which self tensions the taper region and allows the fiber to be
brought into close proximity with the sample surface.  We then dimple the fiber by using 
a bare stripped fiber as a mold, and applying pressure to the narrowest part of the taper 
while heating the contact region~\cite{Michael_07}.  
The dimple, which is typically about 10 $\mu$m in depth, as shown in Fig.~\ref{fig:fiber},  
creates a local evanescent probe to the nanocavity of interest.  The tapered fiber is spliced into 
an optical set-up, and its location with respect to the sample is precisely positioned 
using motorized stages with 50 nm encoder resolution (Surugu Seiki). The photodiode signal 
(Thorlabs Det010FC with a 1 k$\Omega$ load resistor) is passed through a low noise Stanford 
pre-amplifier and low-pass filtered at 1 kHz before computer acquisition.  

The second spectroscopy
technique we use is a cross-polarized resonant scattering set-up, in which light incident from 
normal to the plane of the sample is strongly focused by a 100$\times$ objective and the reflected
signal detected in the cross polarization~\cite{McCutcheon_05}.  
This method allows us to confirm the polarization of
the modes detected by the tapered fiber, since it is only sensitive to the TE modes. 

The cavities were fabricated with standard e-beam lithography and reactive ion 
etching methods, as described previously~\cite{Deotare_09}.

\begin{figure}[t]
\begin{center}
\includegraphics[width=8.6cm]{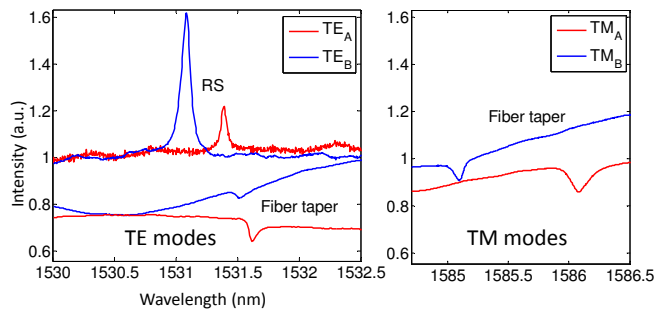}
\end{center}
\vspace{-10pt}
\caption{Spectra from two cavities, A and B, each of which supports two high $Q$-factor modes 
with TE (Q $\sim$ 28,000) and TM (Q $\sim$ 10,000) polarizations.  The fiber taper spectra 
reveal both features, whereas the resonant scattering (RS) spectra resolve only the 
TE modes, since they couple to radiation normal to the device plane.\label{fig:dual}}
\end{figure}

Figure~\ref{fig:dual} shows spectra from two different dual-polarization cavities.  This
represents, to our knowledge, the first experimental demonstration of high-$Q$, TE-TM 
nanocavities.  The cavities are
nominally identical except that the filling fraction of the air holes of cavity A (red) is 
slightly smaller compared to cavity B.  These cavities are not actually fabricated according
to the optimal design, but are intentionally detuned in order to lower the $Q$-factor of the 
TE mode and increase its visibility in the spectra.  This is achieved by reducing the gap
between the two central holes by 20 nm (see Fig.~\ref{fig:modes}).  We used a similar detuning 
method in previous work~\cite{Deotare_09} to predictably shift the $Q$-factor and operating 
wavelength of our nanobeam cavities. Using SEM images of the fabricated structure, we perform 
3D-FDTD simulations of cavity A, and predict $Q_{\rm TE}$ = 27,000 and $Q_{\rm TM}$ = 40,000.

\begin{figure}[b]
\begin{center}
\includegraphics[width=7cm]{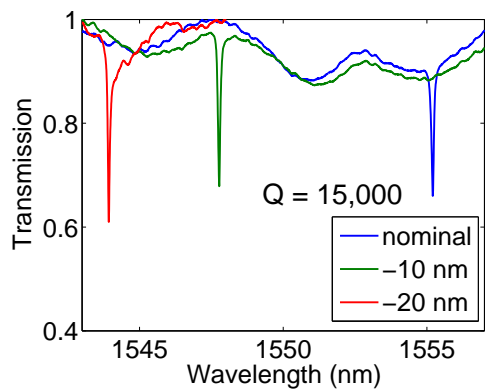}
\end{center}
\vspace{-10pt}
\caption{(a) Fiber spectra of the TM modes from three cavities.  The spacing of the 
central 2 holes of the cavity (as indicated) varies the resonant wavelength, but has
little effect on the $Q$-factor of 15,000.
\label{fig:TM}}
\end{figure}

Measurements from both the fiber taper and resonant scattering set-ups are superimposed
in the same graph.  The TE modes near 1531 nm are revealed using both methods, and have
$Q$-factors of $\sim$ 28,000, as determined by resonant scattering, which is a
non-perturbative technique (i.e. it does not load the cavity).  The
TM modes near 1585 nm show signatures only in the fiber spectra, and have loaded $Q$-factors 
of $\sim$ 10,000.  These measurements compare favorably with the 3D-FDTD results, 
considering the uncertainty in how exactly the fiber loads the cavity (the loaded TM $Q$-factor 
was simulated to be 20,000).
The fiber transmission spectra are acquired by touching the
fiber to the cavity (as visible in the optical image in Fig.~\ref{fig:fiber}), 
which was found to provide greater
stability and repeatibility compared to evanescent coupling from the air.  
In the fiber data, the signatures of both the TE and TM modes
are revealed as dips, since light which is ``dropped'' from the fiber into the cavity 
couples back into the fiber and interferes with the light transmitted directly through the fiber.
On the other hand, in the resonant scattering spectra, free-space light incident from the normal 
direction can only resonantly scatter into the TE modes.  No resonant features are seen at the 
TM wavelength.  The TE modes are revealed 
as peaks on the non-resonant background (although both dips and Fano features~\cite{Galli_09} 
are possible).  
The Lorentzian lineshapes are centered at the bare (unloaded) cavity resonance, and their 
widths give the unloaded $Q$-factors of the cavities.  The fiber spectra are slightly 
red-shifted due to the perturbative effects of the silica fiber, and the exact positions of
the resonances are dependent on how exactly the fiber contacts the particular cavity.

Having demonstrated high-$Q$ TE-TM nanocavities, we now analyze the TM modes in greater detail.  
Although the strength of the fiber coupling to both the TE and TM 
modes is similar in Fig. 1, as measured by the depth of the transmission drop, in general we observe 
a greatly enhanced coupling to the TM modes compared to the TE modes in our fabricated structures.    
In Figure~\ref{fig:TM}, we show a series of TM spectra from three cavities, which differ only in the 
spacing between the central two holes of the cavity.  Each spectrum shows a large transmission 
drop on resonance.  The cavity resonance blue-shifts as the central cavity gap is decreased, 
as expected.  The $Q$-factor shows little variation over this range, and is approximately 
15,000 for each of the cavities.  

\begin{figure}[t]
\begin{center}
\includegraphics[width=7cm]{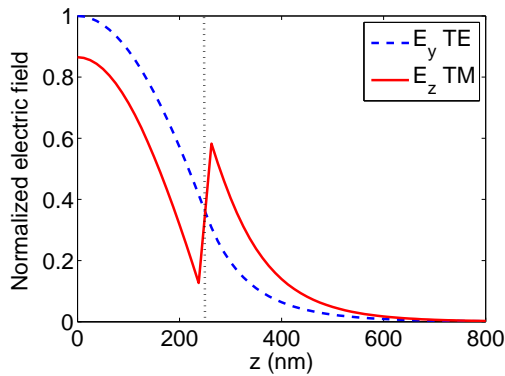}
\end{center}
\vspace{-10pt}
\caption{Comparison between the dominant electric field components of the two modes 
(normalized to the same energy) as a function of distance perpendicular to the device 
plane (z=0 denotes the middle of the cavity, and the dashed line marks the top surface).
\label{fig:coupling}}
\end{figure}

We attribute the greater fiber coupling of the TM modes to their greater overlap with the
fiber mode.  In Fig.~\ref{fig:coupling}, we plot
the electric field strength for both TE and TM modes as a function of $z$ from the middle 
of the cavity.  In the
evanescent region (to the right of the dashed line), the TM E-field is twice the magnitude of the TE
E-field.  This is a consequence of the field boundary conditions.  Since the
electric field of the TM mode is dominantly perpendicular to the silicon/fiber interface, it must
satisfy the boundary condition $\epsilon_1 E_{z,1} = \epsilon_2 E_{z,2}$, where 1 denotes the silicon
and 2 the fiber.  By contrast, the TE mode fields are oriented parallel to the interface, and must
therefore be continuous across the boundary i.e. $E_{y,1}=E_{y,2}$.  The on-resonance
transmission can be estimated using coupled mode theory~\cite{Spillane_03,Barclay_05}, where
the relevant coupling rates are extracted from FDTD simulations.  The TM mode was designed to 
have a waveguide-limited $Q$-factor of 120,000.  With the fiber touching the nanobeam, the 
loaded cavity $Q$-factor is 20,000, and the waveguide $Q$ of the cavity is 180,000, from which
we estimate a transmission of 0.7, a value consistent with our data (Fig.~\ref{fig:TM}).

In conclusion, using a combination of fiber taper and resonant scattering spectroscopy, we 
have experimentally demonstrated high $Q$-factor, dual-polarized TE-TM photonic crystal 
nanobeam cavities in silicon.  
The modes are separated by 50 nm, and each has a $Q$-factor greater than $10^4$.  
We observe large coupling of the TM modes in fiber taper transmission measurements, which
we attribute to the significant evanescent tail of the TM mode above the cavity.
We anticipate this phenomenon could be exploited for 
certain applications, such as bio-sensing~\cite{Vollmer02}.  More generally, we foresee 
TE-TM nanocavities playing an enabling role in many novel integrated applications, such 
as nonlinear wavelength conversion and quantum-cascade lasers.


\vspace{-20pt}

\end{document}